\documentclass[aps,prd,twocolumn,amssymb,amsmath,floatfix,nofootinbib,superscriptaddress,preprintnumbers]{revtex4-1}

\usepackage{graphicx}
\usepackage{dcolumn}
\usepackage{bm}
\usepackage[usenames, dvipsnames]{color}
\usepackage[toc,page]{appendix}
\usepackage[normalem]{ulem}
\usepackage{sidecap}
\usepackage{subcaption}
\usepackage{tabularx,ragged2e,pict2e,booktabs}
\usepackage[colorlinks,bookmarks]{hyperref}
\usepackage{newtxtext,newtxmath}
\usepackage[T1]{fontenc}
\DeclareRobustCommand{\VAN}[3]{#2}
\let\VANthebibliography\thebibliography
\def\thebibliography{\DeclareRobustCommand{\VAN}[3]{##3}\VANthebibliography}
\usepackage{graphicx}	
\usepackage{amsmath}	
\definecolor{linkblue}{rgb}{0,0,0.8}
\definecolor{linkgreen}{rgb}{0,0.5,0}

\hypersetup{pdfpagemode=UseNone, pdfstartview=FitH, linkcolor=linkblue, %
            citecolor=linkblue, urlcolor=linkblue}

\graphicspath{{Figures/}}

\usepackage{soul}
\setstcolor{red}

\begin{document}
\title[Constraints on the scalar-field potential in warm inflation]{Constraints on the scalar-field potential in warm inflation}
\author{Gabriele Montefalcone}
\email{montefalcone@utexas.edu}
\affiliation{Center for Cosmology and Astroparticle Physics, Weinberg Institute for Theoretical Physics, Department of Physics, University of Texas, Austin, Texas 78751, USA}

\author{Vikas Aragam}
\affiliation{Center for Cosmology and Astroparticle Physics, Weinberg Institute for Theoretical Physics, Department of Physics, University of Texas, Austin, Texas 78751, USA}

\author{Luca Visinelli}
\affiliation{Tsung-Dao Lee Institute (TDLI),
520 Shengrong Road, 201210 Shanghai, People's Republic of China}
\affiliation{School of Physics and Astronomy, Shanghai Jiao Tong University,
800 Dongchuan Road, 200240 Shanghai, People's Republic of China}

\author{Katherine Freese}
\affiliation{Center for Cosmology and Astroparticle Physics, Weinberg Institute for Theoretical Physics, Department of Physics, University of Texas, Austin, Texas 78751, USA}
\affiliation{The Oskar Klein Centre, Department of Physics, Stockholm University, AlbaNova, SE-10691 Stockholm, Sweden}
\affiliation{Nordic Institute for Theoretical Physics (NORDITA), 106 91 Stockholm, Sweden
}

\begin{abstract}
We quantify the degree of fine tuning required to achieve an observationally viable period of inflation in the strongly dissipative regime of warm inflation. The ``fine-tuning'' parameter $\lambda$ is taken to be the ratio of the change in the height of the potential $\Delta V$ to the change in the scalar field $(\Delta \phi)^{4}$, i.e. the width of the potential, and therefore measures the requisite degree of flatness in the potential. The best motivated warm inflationary scenarios involve a dissipation rate of the kind $\Gamma\propto T^c$ with $c\geq 0$, and for all such cases, the bounds on $\lambda$ are tighter than those for standard cold inflation by at least 3 orders of magnitude. In other words, these models require an even flatter potential than standard inflation. On the other hand for the case of warm inflation with $c< 0$, we find that in a strongly dissipative regime the bound on $\lambda$ can significantly weaken with respect to cold inflation. Thus, if a warm inflation model can be constructed in a strongly dissipative, negatively temperature-dependent regime, it accommodates steeper potentials otherwise ruled out in standard inflation.
\end{abstract}
\preprint{UTWI-03-2022, NORDITA 2022-068}
\maketitle

\section{Introduction}

Inflation~\cite{Guth:1980zm,Linde:1981mu, Albrecht:1982wi,Kazanas:1980tx, Starobinsky:1980te, Sato:1981ds, Sato:1980yn, Mukhanov:1981xt, Linde:1983gd,Mukhanov:2003xw} is currently the most convincing mechanism to address the horizon, flatness, and monopole problems in the standard Big-Bang cosmology. The accelerated expansion rate during inflation assures that when the process ends the Universe is sufficiently flat, homogeneous and isotropic at the largest observable scales. In addition to solving the problems posed by the Big-Bang cosmological theory, inflation provides a mechanism for generating the fluctuations that seed the inhomogenities we observe in the cosmic microwave background (CMB)~\cite{Planck:2018nkj}. Many models of inflation involve a single scalar field, the inflaton, slowly rolling down a nearly flat potential, inducing a quasi-de Sitter phase. In these models, the density fluctuations are adiabatic and originate from the quantum fluctuations of the inflaton. A universal feature of most potentials implemented in this framework is that they tend to overproduce density fluctuations, unless the potential for the slowly rolling field is chosen very carefully.

The degree of the fine-tuning necessary for a successful  inflationary model that uses a slow-rolling field $\phi$ was studied quantitatively in~\cite{Adams:1990pn}. Specifically, they derived general bounds on a ``fine-tuning'' parameter $\lambda$, defined as:
\begin{equation}
    \lambda \equiv \frac{\Delta V}{(\Delta \phi)^4}, \label{eq:1.1}
\end{equation}
where $\Delta V$ is the decrease in the potential $V(\phi)$ during the inflationary epoch and $\Delta \phi$ is the change in the value of the field $\phi$ over the same period. The parameter $\lambda$ is thus the ratio of the height of the potential to its width, i.e.\ a measure of the degree of flatness of the potential. The authors in~\cite{Adams:1990pn} found that for a standard inflationary model to be observationally consistent, the potential has to be extremely flat, with a fine-tuning parameter $\lambda\lesssim 10^{-6}-10^{-9}$. This bound on $\lambda$ translates into a bound on the quartic coupling constant $\lambda_q$ of the underlying microphysical theory. Given a quartic polynomial potential monotonically decreasing over the interval of interest\footnote{Note, this is a necessary feature of the slowly rolling solution.} we have $|\lambda_q|\leq 36 \lambda$~\cite{Adams:1990pn}, where the quartic term in the Lagrangian is written as $\frac{1}{4}\lambda_q\phi^4$. Note that the numerical values of the bounds above were obtained for the 8 $e$-folds of inflation during which density perturbations are produced on observable scales, so that for typical potentials the bounds over a longer period of inflation are significantly tighter, e.g.\ for 60 $e$-folds $\lambda_q\lesssim 10^{-12}$.

Although the inflaton is often taken to be only (minimally) coupled to gravity, introducing couplings to other early Universe sectors can relax various restrictions normally imposed in inflationary models. A well-established alternative framework to conventional (cold) inflation is warm inflation, in which the inflaton is thermally coupled to a bath of radiation~\cite{Berera:1995ie,Berera:1995wh, Berera:1996fm}. Fluctuations in warm inflation are predominantly thermal in origin, with quantum fluctuations being subdominant in the limit of a large dissipation rate between the two sectors. Additionally, the inflaton continually sources the production of radiation, which alleviates the need for a separate reheating phase at the end of inflation.\footnote{For a recent model of warm inflation with a double scalar field, see ~\cite{DAgostino:2021vvv}.}

In this paper, we investigate for the first time the bound on the parameter $\lambda$ arising in the strong dissipative regime of warm inflation. We find that the friction induced by the dissipation reduces the required width of the potential for a given number of $e$-folds of inflation, while the size of perturbations is increased for a non-negative temperature positive dependence of the friction term, leading to an overall stringent requirement on the fine-tuned potential. In Sec.~\ref{sec:2}, we review the general properties and predictions of warm inflation. In Sec.~\ref{sec:3}, we formulate the problem for strongly dissipative inflationary models, in analogy with what was done for standard inflationary models in~\cite{Adams:1990pn}. In Sec.~\ref{sec:4}, we complete the derivation and find bounds on the fine-tuning parameter $\lambda$ both for the general case and the special cases of a constant Hubble parameter and dissipation strength. Finally, we conclude in Sec.~\ref{sec:5} with a summary of our results. We work in natural units with $c = \hbar = k_B = 1$.

\section{Background on Warm Inflation}\label{sec:2}

In the warm inflation scenario, a substantial fraction of the inflaton energy is converted into radiation during the inflationary period. This mechanism is parameterized by the introduction of a non-negligible dissipation (source) rate $\Gamma$ in the dynamics of the inflaton field (radiation density): 
\begin{align}
    &\Ddot{\phi}+(3H+\Gamma)\dot{\phi}+V_{,\phi}=0\,,\label{eq:2.1} \\
    &\Dot{\rho}_R+4H\rho_R=\Gamma \Dot{\phi}^2\,,  \label{eq:2.2}
\end{align}
where a dot denotes the derivative with respect to cosmic time and $V_{,\phi} \equiv {\rm d} V/{\rm d} \phi$. The criterion for warm inflation to occur is that the thermal fluctuations dominate over the quantum fluctuations, which simply amounts to $H<T$~\cite{Berera:1995ie}.

Inflation is realized when the Hubble expansion rate $H$ is approximately constant. This is achieved when the potential $V(\phi)$ dominates over all other forms of energy. In terms of the background evolution this amounts to:
\begin{equation}
    H^2\simeq \frac{V}{3M_{\mathrm{pl}}^2}, \label{eq:2.3}
\end{equation}
where $H\equiv \dot{a}/a$ is the Hubble parameter, $a$ is the scale factor of the Universe and $M_{\mathrm{pl}}= 1/\sqrt{8 \pi G} \, \approx 2.436\times 10^{18}\,\mathrm{GeV}$ is the reduced Planck mass. During this period, known as the slow-roll regime of the inflaton field, higher order derivatives in Eqs.~\ref{eq:2.1} and~\ref{eq:2.2} can be neglected,
\begin{equation}
    \ddot{\phi} \ll H \dot{\phi}, \quad \text { and } \quad \dot{\rho}_{R} \ll H \rho_{R}\,. \label{eq:2.4}
\end{equation}
As a result, in this regime the equation of motion for the inflaton and the radiation bath respectively read:
\begin{align}
 \Dot{\phi}\simeq -\frac{V_{,\phi}}{3H(1+Q)}\,, \quad \rho_R\simeq \frac{3Q \Dot{\phi}^2}{4}\,, \label{eq:2.5}
\end{align}
where the dissipation strength 
\begin{equation}
Q\equiv \Gamma/(3H)
\end{equation}
is a dimensionless ratio that measures the effectiveness at which the inflaton converts into radiation. Additionally, we assume that the radiation thermalizes on a time scale much shorter than $1/\Gamma$~\cite{Berera:1995ie, Berera:1995wh}, so that the energy density of radiation can be taken to be:
\begin{equation}
    \rho_R(T)=\alpha_1 T^4, \quad \text{with } \quad \alpha_1=\frac{\pi^2}{30}g_{*}(T), \label{eq:2.6}
\end{equation}
where $g_{*}(T)$ is the number of relativistic degrees of freedom of radiation at temperature $T$. Combining Eq.~\ref{eq:2.6} with Eq.~\ref{eq:2.5} gives us the evolution of the temperature of the radiation bath during the inflationary period:
\begin{equation}
    T\simeq \left( \frac{3Q \Dot{\phi}^2}{4\alpha_1}\right)^\frac{1}{4}.
    \label{eq:2.1.8}
\end{equation}

The friction terms present in Eq.~\ref{eq:2.5} modify the standard Hubble slow-roll parameter, which now amounts to:
\begin{equation}
    \epsilon_H\equiv -\frac{\dot{H}}{H^2}\simeq \frac{\epsilon_V}{1+Q}= \frac{M_{\mathrm{pl}}^{2}}{2(1+Q)}\left(\frac{V_{, \phi}}{V}\right)^{2}.
\end{equation}
The accelerated inflationary period terminates when $\epsilon_H = 1$, i.e.\ when $\epsilon_V = 1 + Q$. This can be thought of as a generalization of the slow-roll conditions obtained in cold inflation, for which we would set $Q=0$. For $Q\gg 1$, the slow-roll conditions are substantially relaxed and can in principle be satisfied by scalar field potentials that would otherwise violate the standard slow-roll conditions in the cold inflation scenario.

The presence of a radiation bath and a dissipation rate not only alters the background dynamics of the inflaton field but also its perturbations. Specifically, in the strong dissipative regime ($Q\gg1$), the thermal inflaton perturbations dominate over the usually considered quantum 
fluctuations and the primordial power spectrum takes the general form~\cite{Hall:2003zp,Bastero-Gil:2011rva,Graham:2009bf,Ramos:2013nsa}:
\begin{equation}
    \Delta_{\mathcal{R}}^{2}\simeq\left(\frac{H^{2}}{2 \pi \dot{\phi}}\right)^{2}\sqrt{3\pi Q} \left(\frac{T}{H}\right) G(Q).
    \label{eq:2.1.9}
\end{equation}
Here, the function $G(Q)$ accounts for the coupling of the inflaton and radiation fluctuations due to a temperature-dependent dissipative coefficient, 
\begin{equation}
    \Gamma\propto T^c\,,
\end{equation}
with $c\neq0$. The function $G(Q)$ can only be determined numerically by solving the full set of perturbation equations. Thus, $G(Q)=1$ if $c=0$, while for $c\neq0$ and $Q\gg1$, we can approximate $G(Q)$ with a power-law function of $Q$, i.e. 
\begin{equation}
G(Q)\simeq a_G Q^{b_G} \, ,
\end{equation}
where the prefactor is generally $a_G\sim \mathcal{O}(1)$ and the exponent is $b_G>0$ $(<0)$ for a positive (negative) temperature dependence. For $c\geq 0$, the amplitude of the scalar perturbations in warm inflation is enhanced compared to standard cold inflation; the larger the power $c>0$, the larger the enhancement as the fluctuations get coupled earlier~\cite{Bastero-Gil:2011rva}. For $c<0$ we have the opposite effect, and
the spectrum is diminished with respect to the $c=0$ case and can be lower than the standard cool inflation prediction in an extreme dissipative regime.

In this Letter, we take $Q \gg 1$, describing the limit of a strong thermal dissipation. We focus on the strongly dissipative regime since it is only in this extreme scenario that we can expect significant changes in both the numerator and the denominator of $\lambda$ in Eq.~\ref{eq:1.1}, relative to cold inflation. Specifically, the case $Q\gg 1$ has two distinct features: i) the function $G(Q)$ in Eq.~\ref{eq:2.1.9} is substantially greater than one, leading to a modification of the possible values for the Hubble rate and the field velocity $\dot\phi$ once the constraint on the size of density perturbations (see Eq.~\ref{eq:3.1}) is satisfied; ii) the condition on the slow-roll parameter is significantly looser which allows the field excursion $\Delta \phi$ to be much smaller than $M_{\rm Pl}$ as we will see in more detail in Sec.~\ref{sec:4.0}.

\section{Formulation of the problem for warm inflationary models}
\label{sec:3}

The derivation of the bounds on the fine-tuning parameter $\lambda$ presented in~\cite{Adams:1990pn} essentially boils down to two effects: (i) the width of the potential must allow a sufficiently large number of $e$-folds during inflation and (ii)
the height of the potential is constrained by observational constraint on the density perturbations $\delta\rho/\rho$, together with the so called overdamping constraint, that arises from the consistency of neglecting the $\ddot{\phi}$ term in Eq.~\ref{eq:2.1}.  Note that \cite{Adams:1990pn} restrict their discussion to the 8 $e$-folds of inflation relevant to the production of perturbations observable in the CMB, i.e.\ on scales of size $1-1000\,$Mpc. 

Here, we follow the same prescription and take the density perturbations constraint to be:\footnote{Note that $\delta = A_s^{1/2}$, where $A_s$ is the amplitude of curvature perturbations as measured by the {\it Planck} Collaboration~\cite{Planck:2018nkj}.}
\begin{equation}
    \frac{\delta \rho}{\rho}\Big|_{\mathrm{\mathrm{cmb}}}\equiv \Delta_{\mathcal{R}}\Big|_{\mathrm{cmb}}\leq \delta\approx 5\times 10^{-5}, \label{eq:3.1}
\end{equation}
which can be rewritten by combining Eqs.~\ref{eq:2.1.8},~\ref{eq:2.1.9} and~\ref{eq:3.1} in the more convenient form:
\begin{equation}
    \left(\frac{H^2}{\Dot{\phi}}\right)\lesssim \left[\frac{(2\pi)^8}{\alpha_s}\right]^{\frac{1}{6}}\frac{\delta^{\frac{4}{3}}}{Q^{\frac{1}{2}+\frac{2}{3}b_G}}, \label{eq:3.3}
\end{equation}
with $\alpha_s\equiv 27\pi^2a_G^4/(4\alpha_1)$. This clearly differs from the equivalent constraint for standard cold inflation, i.e.\ $(H^2/\Dot{\phi})\lesssim 2\pi\delta$, specifically in the presence of an explicit dependence in $Q$ and by an additional factor $(2\pi\delta)^{\frac{1}{3}}/\alpha_s^{\frac{1}{6}}$.

We further derive the overdamping constraint on the potential to be:
\begin{equation}
    \left|\frac{{\rm d}}{{\rm d} t}\left[\frac{1}{3 H(1+Q)} \frac{{\rm d} V}{{\rm d} \phi}\right]\right| \leq\left|\frac{1}{(1+Q)}\frac{{\rm d} V}{{\rm d} \phi}\right|. \label{eq:3.2}
\end{equation}
Note, Eq.~\ref{eq:3.2} is valid for any value of $Q\geq 0$ and in fact, for $Q=0$ (standard cold inflation) it reduces to the overdamping constraint derived in~\cite{Adams:1990pn}. Since in this Letter we are only interested in the strong dissipative regime ($Q\gg1$), we can always safely take $1+Q\simeq Q$ in Eq.~\ref{eq:3.2}. 

We assume that both constraints in Eqs.~\ref{eq:3.1}--\ref{eq:3.2} hold for the relevant period of $N\approx8$ $e$-foldings that can be probed in the CMB. In addition, we adopt a new time variable $x$ defined in Eq. {\ref{eq:3.4}} in terms of the number $n$ of $e$-foldings since the beginning of the epoch,
and in this new notation rewrite the overdamping 
constraint  as Eq. {\ref{eq:3.5}}
and density perturbation constraint as Eq. {\ref{eq:3.6}}:
\begin{align}
    &{\rm d}x\equiv \frac{{\rm d}n}{N}=\frac{H{\rm d}t}{N}, \label{eq:3.4} \\
    &\left|H \frac{{\rm d}}{{\rm d} x}\left[\frac{F}{QH}\right]\right| \leq \frac{3 N F}{Q}, \label{eq:3.5} \\
    &\frac{3QH^3}{F}\lesssim \left[\frac{(2\pi)^8}{\alpha_s}\right]^{\frac{1}{6}}\frac{\delta^{\frac{4}{3}}}{Q^{\frac{1}{2}+\frac{2}{3}b_G}}, \label{eq:3.6}
\end{align}
where the variable $x$ ranges from 0 to 1 during the
relevant time period and
\begin{equation}
 F(x)\equiv -\frac{dV}{d\phi}. 
\end{equation}
 Finally, we write the quantities $\Delta V$ and $\Delta \phi$ as
\begin{align}
    \Delta V&=\frac{N}{3} \int_{0}^{1}\left(F^{2} / QH^{2}\right) {\rm d} x=\frac{N\Bar{F}^2}{3\Bar{Q}\Bar{H}^2}\int^1_0\frac{f^2}{qh^2}{\rm d} x, \label{eq:3.7}\\
    \Delta \phi&=\frac{N}{3} \int_{0}^{1}\left(F / QH^{2}\right) {\rm d} x =\frac{N\bar{F}}{3\Bar{Q}\Bar{H}^2}\int^1_0\frac{f}{qh^2}{\rm d} x, \label{eq:3.8}
\end{align}
where we introduced the dimensionless functions:
\begin{equation}
    f(x)\equiv F(x)/\Bar{F}, \quad h(x)\equiv H(x)/\Bar{H}, \quad q(x)\equiv Q(x)/\Bar{Q}. \label{eq:3.9}
\end{equation}
Here, the bar refers to the value of the functions evaluated at $x=\Bar{x}$, which denotes the value of $x$ such that the quantity $QH^3/F$, that appears in the density perturbation constraint in Eq. {\ref{eq:3.6}}, is maximized.\footnote{If the maximum is not unique, we can choose $x$ to be any of the maxima.} It is useful to note that in Eqs.~\ref{eq:3.7}--\ref{eq:3.8} we chose the same sign convention as in~\cite{Adams:1990pn} such that $\Delta V$ is a positive quantity and $x=0$ at the beginning of the constrained time period.

Using the definition of $\lambda$ in Eq.~\ref{eq:1.1} and Eqs.~\ref{eq:3.7}-\ref{eq:3.8}, we can write the fine-tuning parameter as:
\begin{equation}
\lambda=\frac{3\bar{Q}}{N^{3}}\left[\frac{3 \bar{Q}\bar{H}^{3}}{\bar{F}}\right]^{2} J[f,q,h], \label{eq:3.10}
\end{equation}
with:
\begin{equation}
J[f, q,h] \equiv \frac{\int_{0}^{1}\left(f^{2} / qh^{2}\right) {\rm d} x}{\left[\int_{0}^{1}\left(f / qh^{2}\right) {\rm d} x\right]^{4}}. \label{eq:3.11}
\end{equation}
Adding the density perturbation constraint to Eq.~\ref{eq:3.10} we find:
\begin{align}
    \lambda\lesssim \frac{3 J[f,q,h]}{N^3} \left[\frac{(2\pi)^8}{\alpha_s}\right]^{\frac{1}{3}} \frac{\delta^{\frac{8}{3}}}{\bar{Q}^{\frac{4b_G}{3}}}. \label{eq:3.12}
\end{align}
We note that $\lambda$ is proportional to $\delta^{\frac{8}{3}}$ instead of $\delta^2$, as in the cold inflation scenario. Additionally, $\lambda$ has an explicit $Q$ dependence only if $b_G\neq 0$, which is true for a temperature-dependent dissipation rate ($b_G\neq0$ i.f.f.\ $\Gamma\propto T^c$ with $c\neq0$).

In order to derive an upper bound on the fine-tuning
parameter, we must derive an upper bound on the functional
$J[f,q,h]$, subject to the following constraints:
\begin{align}
qh^{3}(x) / f(x) &\leq 1, &  \forall x \in[0,1], \label{eq:3.13} \\
\left|\frac{1}{f} \frac{{\rm d} f}{{\rm d} x}-\left(\frac{1}{h} \frac{{\rm d} h}{{\rm d} x}+\frac{1}{q} \frac{{\rm d} q}{{\rm d} x}\right)\right| &\leq 3 N, &  \forall x \in[0,1]. \label{eq:3.14}
\end{align}
The first one of these inequalities directly follows from the rescaling of the functions relative to the maximum value of $QH^3/F$ and is related to the density perturbation constraint, while the second inequality is the overdamping constraint in Eq.~\ref{eq:3.5}. Finally, we also know that there exists a point $x\in[0, 1]$ such that $f(\bar{x})=h(\bar{x})=q(\bar{x})=1$.

\section{Constraints on warm inflationary models} \label{sec:4}

Using the formulation of the problem given above,
we now present the constraints on the fine-tuning parameter $\lambda$ in warm inflationary models. We begin in Sec.~\ref{sec:4.0} with the simplified case of a constant Hubble parameter $H$, dissipation strength $Q$, and slope of the potential $V_\phi$. Using the bounds on the field excursion as well as the size of the density fluctuations, this simple argument illustrates the basic power-law dependence of the fine tuning-tuning parameter $\lambda$ on the dissipation strength $Q$. Next, in Sec.~\ref{sec:4.1} we turn to the general case, in which $f$, $h$ and $q$ are all arbitrary functions to be chosen independently. Finally, in Sec.~\ref{sec:4.2} we consider the special case of a constant Hubble parameter ($h =1$), for which a stronger bound on $\lambda$ than in the general case can be derived. For each of the cases we study, we also quote the corresponding bound obtained in \cite{Adams:1990pn} for the case of cold inflation.

In order to obtain numerical values for our limits, in the following we consider the representative case in which $\delta = 5\times 10^{-5}$, $N=8$, $a_G=1$ and $g_{*}(T)=228.75$ (corresponding to the number of relativistic degrees of freedom in the minimal supersymmetric Standard Model). 

\subsection{An intuitive picture}
\label{sec:4.0}

We begin with a simple argument for the bounds on the fine-tuning parameter $\lambda$ in the case of a constant Hubble parameter $H$ and dissipation strength $Q$, assuming a linear potential throughout inflation. For this fiducial case, we can compute directly the field excursion $\Delta\phi$ and the potential change $\Delta V$ by saturating the slow-roll condition, i.e.\ $\epsilon_V\leq Q$, in combination with the bound on the observations on the scale of the fluctuations. 

 The field excursion $\Delta\phi$ in a strongly dissipative regime can be significantly reduced compared to the case of no dissipation for the same number of $e$-folds. This is because the decay of the inflaton into radiation is effectively playing the role of an additional friction term on top of the usual Hubble friction from standard cold inflation. In formulas, we get:
\begin{equation}
    \frac{\Delta \phi}{M_{\mathrm{Pl}}}\approx\frac{\dot{\phi}}{H} N= \frac{\sqrt{2 \epsilon_V}}{Q} N\lesssim \sqrt{\frac{2}{Q}}N, \label{eq:4.0.1}
\end{equation}
where $N$ is the number of \textit{e}-folds and we used the relation ${\rm d}N=H{\rm d}t$, and  the slow-roll condition $\epsilon_V\leq Q$.

In a similar fashion, we can apply the slow-roll condition on $\epsilon_V$ on the potential change $\Delta V$ during the inflationary period, defined according to:
\begin{equation}
    \Delta V \approx \frac{V_\phi \dot{\phi}}{H}N
     = 2N V\frac{\epsilon_V}{Q} \lesssim 2N V. \label{eq:4.0.2}
\end{equation}
These bounds account for the dynamics of the inflaton during slow-roll and need to be combined with the observations on the scale of the fluctuations in Eq.~\ref{eq:3.3} to fix the size of the potential and obtain a bound on $\lambda$. In fact, it turns out that the constraint in Eq.~\ref{eq:3.3} can be recast as an upper bound on the scale of the inflaton $V$, and thus on $\Delta V$ via Eq.~\ref{eq:4.0.2}, by including the slow-roll condition on $\epsilon_V$.\footnote{To accomplish this, we first rewrite Eq.~\ref{eq:3.3} in terms of $V$, $\epsilon_V$ and $Q$, using Eqs.~\ref{eq:2.3} and~\ref{eq:2.5}, and then plug in the slow-roll condition $\epsilon_V\leq Q$ and Eq.~\ref{eq:4.0.2}.} In formulas, we get:
\begin{equation}
    \frac{\Delta V}{M_{\mathrm{pl}}^4}\lesssim 12N\left[\frac{(2\pi)^8}{\alpha_s}\right]^{\frac{1}{3}}\delta^{\frac{8}{3}}Q^{-2-\frac{4}{3}b_G}. \label{eq:4.0.3}
\end{equation}
The corresponding bound in cold inflation reads: $\Delta V/M_{\mathrm{pl}}^4\lesssim 12 N\pi^2\delta^2 $. Thus, for a non-negative temperature dependence on $\Gamma$ ($b_G\geq 0$), the constraint on $\Delta V$ in Eq.\ref{eq:4.0.3} is clearly more stringent than in cold inflation due to its negative dependence on $Q$ as well as higher power of dependence on $\delta$ (which is a small number). The more stringent bound we find here is related to the fact that, in order to reproduce the observed density perturbations, the scale of inflation is reduced to counteract the large thermal enhancement factor in the power spectrum (see Eq.~\ref{eq:2.1.9}). The opposite effect occurs when $\Gamma$ possesses a negative temperature dependence, as the power spectrum is diminished compared to the case $c=0$, so that larger values of $\Delta V$ can in principle be achieved.

We finally combine Eqs.~\ref{eq:4.0.1} and~\ref{eq:4.0.3} to obtain the constraint on the fine tuning parameter $\lambda$ in Eq.~\ref{eq:1.1}, which reads:
\begin{equation}
    \lambda\lesssim \frac{3}{N^3}\left[\frac{(2\pi)^8}{\alpha_s}\right]^{\frac{1}{3}}\delta^{\frac{8}{3}}Q^{-\frac{4}{3}b_G}\approx 2.8\times 10^{-12} Q^{-\frac{4}{3}b_G}. \label{eq:4.0.4}
\end{equation}
As expected, this is equivalent to substituting $J=1$ in Eq.~\ref{eq:3.12}, i.e.\ setting $q=f=h=1$ $\forall x \in [0,1]$. 

For a temperature-independent dissipation rate for which $b_G=0$, the bound on $\lambda$ becomes independent of the dissipation strength $Q$ and amounts to $\lambda\lesssim 2.8\times 10^{-12}$. The corresponding constraint on the fine tuning parameter for standard cold inflation is much looser: $\lambda\lesssim 1.8\times 10^{-9}$~\cite{Adams:1990pn}. This discrepancy gets even larger if we consider a dissipation rate with a positive temperature dependence, i.e.\ $b_G>0$. 
On the other hand, if we choose a dissipation rate with a negative temperature dependence, i.e.\ $b_G<0$, then for large enough values of the dissipation strength (i.e. $Q^{-4 b_G/3}\gtrsim 500$) the bound on $\lambda$ can be weaker than for cold inflation.

Overall, the numerical values obtained for the bound and its power-law dependence on $Q$ match the results from the detailed computation of the following sections. The argument presented here emphasizes the non-trivial interplay of the constraints on the field excursion in Eq.~\ref{eq:4.0.1} and the scale of inflation in Eq.~\ref{eq:4.0.3}. In short, in a strongly dissipative regime, the value of $\Delta\phi$ can be much smaller than in usual cold inflation due to the effects of friction, e.g. $\Delta\phi$ can lie significantly below the Planck scale. At the same time, for most warm inflationary models (for which $\Gamma\propto T^c$ and $c\geq 0$), the scale of inflation $\Delta V$ is also significantly reduced in order to reproduce the observed density perturbations, so that the corresponding bound on $\lambda$ becomes more stringent compared to standard cold inflation.

\subsection{The general case \label{sec:4.1}}

Along with the bound on the fine-tuning parameter $\lambda$ in Eq.~\ref{eq:3.12}, we derive an upper bound for the functional $J[f,q,h]$ in Eq.~\ref{eq:3.11}, subject to the constraints in Eqs.~\ref{eq:3.13} and~\ref{eq:3.14}. For this purpose, it is convenient to define $f_q\equiv f/q$ in terms of which Eq.~\ref{eq:3.11} reads:
\begin{align}
    J[f_q, q,h]&= \frac{\int_{0}^{1}q\left(f_q^{2} / h^{2}\right) {\rm d} x}{\left[\int_{0}^{1}\left(f_q/ h^{2}\right) {\rm d} x\right]^{4}}, \nonumber \\
    &< \frac{\int_{0}^{1}\left(f_q^{2} / h^{2}\right) {\rm d} x}{\left[\int_{0}^{1}\left(f_q / h^{2}\right) {\rm d} x\right]^{4}}\equiv J[f_q,h], \label{eq:4.1.1}
\end{align}
where the first inequality stems from the fact that $q$ is a positive definite function $\leq 1$ $\forall x \in[0,1]$. From Eq.~\ref{eq:4.1.1} it follows that to derive an upper bound on the fine-tuning parameter $\lambda$ we simply need to find an upper bound on $J[f_q,h]$ subject to the following constraints:
\begin{align}
    h^3(x)/f_q(x)&\leq 1 \quad \forall x \in[0,1], \label{eq:4.1.2}\\
\left|\frac{1}{f_q} \frac{{\rm d} f_q}{{\rm d} x} -\frac{1}{h}\frac{{\rm d} h}{{\rm d} x}\right| &\leq 3 N\quad \forall x \in[0,1]. \label{eq:4.1.3}
\end{align}
We now note that the above constraints and the definition of $J[f_q,h]$ are equivalent to those defined for the standard cold inflationary scenario in~\cite{Adams:1990pn}.\footnote{Eqs. 2.8, 2.10a, 2.10b, in their manuscript.} Thus, we can simply quote here the general bound on the functional $J$ derived in~\cite{Adams:1990pn}, which states:
\begin{align} J[f_q,h]&\leq \frac{27N^3}{32}. \label{eq:4.1.4}
\end{align}
Finally, by combining Eq.~\ref{eq:4.1.1} and~\ref{eq:4.1.4} with Eq.~\ref{eq:3.12} yields:
\begin{align}
    \lambda &< \frac{81}{32}\left[\frac{(2\pi)^8}{\alpha_s}\right]^{\frac{1}{3}}\delta^{\frac{8}{3}} \bar{Q}^{-\frac{4}{3}b_G}\approx 1.2\times 10^{-9} \bar{Q}^{-\frac{4b_G}{3}}. \label{eq:4.1.5}
\end{align}
As we anticipated in Sec.~\ref{sec:4.0}, the explicit $Q$ dependence in Eqs.~\ref{eq:4.0.4} and~\ref{eq:4.1.4} is the same. Therefore, the implications of this bound for different dissipation rates, i.e.\ different values of $b_G$, are similar to those presented in the previous section; i.e.\ the bound is tighter than cold inflation for $c\geq0$ and can be weaker for $c<0$.

For this general case, the constraint on the fine tuning parameter for standard cold inflation is $\lambda\leq 6.3\times 10^{-7}$~\cite{Adams:1990pn}, which is at least 3 orders of magnitude looser than the bound obtained here for strongly dissipative warm inflation with $c\geq 0$. For instance for $b_G=0$, we have $\lambda<1.2\times 10^{-9}$, while if we set the dimensionless dissipation strength to the nominal value $\bar{Q}=100$ and $b_G=2.315$,\footnote{This value of $b_G$ corresponds to the case $c=1$ as found in~\cite{Bastero-Gil:2016qru,Benetti:2016jhf}.} we obtain $\lambda<8.1\times 10^{-16}$ which is about 9 order of magnitude smaller than the corresponding bound obtained for the standard cold inflationary scenario in~\cite{Adams:1990pn}. If we take instead $b_G=-1.41$,\footnote{This value of $b_G$ corresponds to the case $c=-1$ as found in~\cite{Motaharfar:2018zyb}.} and keep $\bar{Q}=100$ we find $\lambda<6.9\times 10^{-6}$ which is about 1 order of magnitude weaker than the corresponding cold inflation bound. We also note that for the same $b_G=-1.41$ but $\bar{Q}=20$, the bound on the fine-tuning parameter is $\lambda<3.4\times 10^{-7}$, which is still marginally tighter than the corresponding cold inflation case. This emphasizes that for $c<0$ the bound on $\lambda$ can loosen relative to cold inflation when the dissipation strength is large enough to satisfy $\bar{Q}^{-4 b_G/3}\gtrsim 500$.

Finally, we note that the scenario in which the dissipation strength is constant during the inflationary epoch, i.e\ $q=1$ $\forall x\in[0,1]$, is encoded already in the constraint on $\lambda$ in Eq.~\ref{eq:4.1.5}. In fact, this special case simply amounts to taking $f_q\rightarrow f$ in Eq.~\ref{eq:4.1.1} which does not change the derived upper bound on the functional $J$. 

\subsection{Constraints with a constant Hubble parameter $H$}
\label{sec:4.2}

We now consider the restricted problem in which the Hubble parameter is constant during the inflationary epoch, i.e.\ $h=1$ $\forall x\in[0,1]$. For this case, the functional $J$ can be written as:
\begin{equation}
    J[f,q] \equiv \frac{\int_{0}^{1}\left (f^{2}/q\right) {\rm d} x}{\left[\int_{0}^{1}\left(f/q \right)  {\rm d} x\right]^{4}}, \label{eq:4.2.1}
\end{equation}
with updated constraints that read:
\begin{align}
    q(x)/f(x)&\leq 1, & \forall x \in[0,1], \label{eq:4.2.2}\\
\left|\frac{1}{f} \frac{{\rm d} f}{{\rm d} x} -\frac{1}{q}\frac{{\rm d}q}{{\rm d}x}\right| &\leq 3 N, & \forall x \in[0,1]. \label{eq:4.2.3}
\end{align}

By making the same convenient substitution $f_q=f/q$ as for the general case, we can rewrite $J$ as:
\begin{align}
    J[f_q,q] &= \frac{\int_{0}^{1} qf_q^{2} {\rm d} x}{\left[\int_{0}^{1} f_q  {\rm d} x\right]^{4}}, \nonumber\\ &< \frac{\int_{0}^{1} f_q^{2} {\rm d} x}{\left[\int_{0}^{1} f_q  {\rm d} x\right]^{4}} \equiv J[f_q]. \label{eq:4.2.4}
\end{align}
In a similar fashion to the general case, the problem now amounts to finding an upper bound  on $J[f_q]$, subject to the following constraints:
\begin{align}
    f_q(x)&\geq 1, & \forall x \in[0,1], \label{eq:4.2.5}\\
\left|\frac{1}{f_q} \frac{{\rm d} f_q}{{\rm d} x}\right| &\leq 3 N, & \forall x \in[0,1]. \label{eq:4.2.6}
\end{align}

We again note that the above constraints and the definition of $J[f_q]$ are equivalent to those defined in~\cite{Adams:1990pn} for the constant Hubble parameter case.\footnote{Eqs. 3.26, 2.10a, 2.10b, in their manuscript.} We can therefore quote the equivalent bound previously derived in~\cite{Adams:1990pn}, namely:
\begin{align}
 J[f_q]\leq 3N. \label{eq:4.2.7}
\end{align}
Finally, we can combine Eqs.~\ref{eq:4.2.4} and~\ref{eq:4.2.7} with Eq.~\ref{eq:3.12} to obtain the desired upper limit on the fine-tuning parameter:
\begin{align}
    \lambda &< \frac{9}{N^2}\left[\frac{(2\pi)^8}{\alpha_s}\right]^{\frac{1}{3}}\delta^{\frac{8}{3}} \bar{Q}^{-\frac{4}{3}b_G}\approx 6.2\times 10^{-11} \bar{Q}^{-\frac{4b_G}{3}}, \label{eq:4.2.8}
\end{align}
We again note that the explicit $Q$ dependence in Eq.~\ref{eq:4.2.8} is the same as in Eqs.~\ref{eq:4.0.4} and~\ref{eq:4.1.5}. The implications of this bound for different dissipation rates are equivalent to those described extensively in the previous sections. The only difference is that in the case of constant Hubble parameter, the bound on $\lambda$ is roughly 2 orders of magnitude more stringent that the constraint obtained in the general case presented above in Sec.~\ref{sec:4.1}. This distinctive feature of the constant Hubble parameter case was also found for the standard cold inflationary scenario in \cite{Adams:1990pn}.

\section{Summary} \label{sec:5}

In this Letter, we have studied the bounds on the fine-tuning parameter $\lambda$ in Eq.~\ref{eq:1.1} for a large class of strongly dissipative warm inflationary models, defined through the temperature dependence of the dissipation rate $\Gamma$ at which the inflaton decays into particles in the radiation bath. The constraints have been derived for the general case in which both the Hubble rate and the dissipation strength evolve with time, as well as for the case in which the Hubble rate is assumed to be a constant. We find that in most cases, the fine-tuning parameter is confined to be smaller than the corresponding cold inflationary models. More precisely, if the dissipation rate $\Gamma$ has a non-negative temperature dependence ($\Gamma \propto T^c$ with $c\geq 0$), the bounds on $\lambda$ are at least 3 orders of magnitude more stringent than those derived in standard cold inflation. Additionally, for a strictly positive coefficient $c>0$, we have an explicit negative dependence of the bound on $\lambda$ on the parameter $Q$, which implies that for large values of the dissipation strength $Q$ our constraints become significantly tighter, i.e.\ $\lambda\lesssim 10^{-15}-10^{-20}$ for $Q\sim 10^{2}-10^{4}$ when $b_G\sim \mathcal{O}(1)$, as in Eq.~\ref{eq:4.1.5}. In the case of a constant value of the Hubble parameter, the constraints on $\lambda$ are around 2 orders of magnitude even more stringent than those obtained in the general case, similarly to what is found for cold inflation. 

Overall, for all the models mentioned above the scalar field potential has to be flatter than what is required for cold inflation in order to build a successful warm inflationary model. The only exception is for models characterized by a dissipation rate with negative temperature dependence $(c<0)$, for which the bound on $\lambda$ shows a positive dependence on $Q$. In this case, at large  values of the dissipation strength the bound on $\lambda$ substantially relaxes, implying that we can accommodate much steeper potentials that are otherwise ruled out in standard cold inflation.

As a whole, these results are counter-intuitive. One would expect the bounds on $\lambda$ to be looser regardless of the form of the dissipation rate $\Gamma$ since, in a strongly dissipative regime, the field excursion $\Delta\phi$ is substantially reduced, see Eq.~\ref{eq:4.0.1}. However, one must also take into account the density perturbation constraint which, for a non-negative temperature dependence on $\Gamma$, sets a much tighter bound on the scale of inflation $\Delta V$ due to the thermal enhancement factor in the amplitude of the scalar perturbations, as in Eq.~\ref{eq:4.0.3}. In other words, unless we significantly lower the energy scale of the inflaton potential compared to a cold inflationary scenario, models of strongly dissipative warm inflation (with $\Gamma\propto T^c$ and $c\geq 0$) generally overproduce density fluctuations. These competing effects on the field excursion and the scale of inflation interact in a non-trivial manner and result in more stringent bounds on $\lambda$ compared to standard cold inflation for most warm inflationary models, except those with a negative temperature dependence on $\Gamma$ (i.e.\ those models which allow a smaller amplitude of the scalar perturbation compared to cold inflation).

So far, most explicit constructions of dissipative terms for warm inflationary models show a positive dependence with temperature. Specifically, a cubic temperature dependence is obtained in the low-temperature regime for warm inflation, in which the inflaton couples only to the heavy intermediate fields whose masses are larger than the radiation temperature~\cite{Berera:2008ar,Bastero-Gil:2010dgy,Bastero-Gil:2012akf,Berghaus:2019whh,Motaharfar:2018zyb}. In contrast, a linear temperature dependence is obtained in the high-temperature regime where the inflaton is directly coupled to the radiation fields and is protected from large thermal corrections due to the symmetries obeyed by the model~\cite{Bastero-Gil:2016qru,Berghaus:2020ekh,Motaharfar:2018zyb}. Also in a high-temperature regime of warm inflation, one can also produce an inversely temperature-dependent dissipation rate ~\cite{Berera:1998gx, Berera:1998px, Motaharfar:2018zyb}; to date the only explicit physical construction of this type was derived in~\cite{Bastero-Gil:2019gao}.\footnote{See also Ref.~\cite{Haddad:2022} for a non-trivial construction.} This further emphasizes the main result of our work. While it is possible to construct a model of warm inflation with relaxed bounds on the fine tuning parameter, for most warm inflationary models of physical interest the requirements on the flatness of the scalar field potential are very stringent and significantly more severe than those found in the cold inflationary scenario.

As a final remark, we emphasize that the more stringent bound demanded on the value of $\lambda$ within the warm inflation framework does not particularly lead to disfavoring any inflation potential model. For example, the class of runaway potentials such as those studied in Ref.~\cite{Das:2020xmh} naturally produces very small values of $\lambda$ which are in agreement with the bounds found in this work. Additionally, models with monomial power law potentials and a positively temperature dependent dissipation coefficient are generally only compatible with observations in the weak dissipative regime $Q \ll 1$~\cite{Benetti:2016jhf}, for which the same bounds on $\lambda$ from cold inflation apply.

\section*{Acknowledgements}

K.F.\ is Jeff \& Gail Kodosky Endowed Chair in Physics at the University of Texas at Austin, and K.F.\ and G.M.\ are grateful for support via this Chair. K.F.\ and G.M.\ acknowledge support by the U.S.\ Department of Energy, Office of Science, Office of High Energy Physics program under Award Number DE-SC-0022021 as well as support from the Swedish Research Council (Contract No.~638-2013-8993). V.A.\ acknowledges support from the National Science Foundation under grant number PHY-1914679.

\bibliographystyle{apsrev4-1}
\bibliography{main}

\end{document}